\documentclass[final,5p,times,twocolumn]{elsarticle}

\usepackage{graphics}
\usepackage{graphicx}
\usepackage{amssymb}
\usepackage{amsthm}
\usepackage{xcolor}
\usepackage{lineno}
\usepackage{lscape}

\biboptions{compress, sort}

\begin{document}

\begin{frontmatter}

\title{Density-Diffusion Relationship in Soda-Lime Phosphosilicate}
\author[a]{Youssef Ouldhnini}
\author[b,c]{Achraf Atila\corref{cor1}}
\ead{achraf.atila@fau.de; achraf.atila@gmail.com}
\ead[url]{www.aatila.com}
\author[d]{Said Ouaskit}
\author[a]{Abdellatif Hasnaoui}
\address[a]{LS2ME, Facult\'{e} Polydisciplinaire Khouribga, Sultan Moulay Slimane University of Beni Mellal, B.P 145, 25000 Khouribga, Morocco}
\address[b]{Department of Materials Science $\&$ Engineering, Institute I: General Materials Properties, Friedrich-Alexander-Universit\"{a}t Erlangen-N\"{u}rnberg, 91058 Erlangen, Germany}
\address[c]{Computational Materials Design, Max-Planck-Institut f\"{u}r Eisenforschung Max-Planck-Str. 1, 40237 D\"{u}sseldorf, Germany}
\address[d]{Laboratoire de Physique de la Matière Condensée, Faculté des Sciences Ben M'sik, University Hassan II of Casablanca, B.P 7955, Av Driss El Harti, Sidi Othmane, Casablanca, Maroc}
\cortext[cor1]{Corresponding author.}

\begin{abstract}
Bioactive glasses release ions such as sodium when implanted in the human body. However, an excess of the released ions can cause problems related to cytotoxicity. The ion release control is considered one of the primary challenges in developing new bioactive glasses. Here, we use molecular dynamics simulations to investigate the effect of the density on atoms' dynamics in an archetypal phosphosilicate bioactive system. The diffusion coefficient displays three main regimes with increasing density. The mobility of the modifiers is significantly affected by the increase of the density, especially Na, compared to other elements. We use a modified Arrhenian model to access the complex dynamic behavior of 45S5 melts and correlate it to the structural changes by evaluating the network connectivity and pair-excess entropy. Overall, our results present a step toward the rational design of bioactive glasses and a key to controlling the ion release of bioactive glasses.
\end{abstract}

\end{frontmatter}

\section{\label{sec:Introduction}Introduction}
Oxide glasses have recently found applications in medicine, particularly in bone implantation engineering or prosthesis \cite{Zhang2009, Best2008}, which is due to osteoconduction and osteoinduction that allow bone growth \cite{Boccaccini2005} and make these bioactive glasses perfect candidates as implants to repair and replace the diseased and damaged bone tissues.
 
In the late sixties, Hench and colleagues made the first bioactive oxide glass \cite{Hench1971}, a soda-lime phosphosilicate glass, 45S5, which is considered as a reference system for bioactive glasses. The 45S5 bioglass can form a strong interfacial bond with soft and hard tissues \cite{Hench1971} and shows antibacterial properties due to the release of alkali and alkaline earth ions \cite{Hu2008, Begum2016, Fiume2019}. The bioactivity and biochemical compatibility of glass are linked generally to its ability to form a hydroxycarbonate apatite (HCA) layer on its surface that contributes to the bonding with surrounded soft and hard tissues \cite{Suchanek1998, Hench2002}. Furthermore, chemical biocompatibility is manifested by the similarity of the glass composition to that of the natural bone by the presence of phosphorus and calcium, and this bioactive glass also enhances healing operation by angiogenesis stimulation ability for \textit{in vivo} bone growth and repair \cite{Gerhardt2011, Handel2013}. 
Three key compositional features make the 45S5 compositions different from the traditional soda-lime silicate glasses: (I) it contains less than 60 mol$\%$ SiO$_{2}$, which is a sufficient amount for the precipitation of the loose silica-rich layer after merging implant in the body fluid, and give network connectivity (NC) around 2 \cite{Gerhardt2010}; (II) it has a high amount of both Na$_{2}$O, and CaO which enhance the homogenization of the glass in the melting process stage \cite{Lin2005}, and to give more strength to the glass network; (III) a high CaO/P$_{2}$O$_{5}$ ratio which enables the glass to enhance the HCA layer formation. This HCA layer is a result of a chemical reactions sequence in the initial dissolution stage when the implant contacts an extracellular fluid \cite{Gerhardt2010}.
When a bioactive glass is in contact with physiological fluid, each element has its specific timing in the initial stage of glass degradation to form the silica-rich layer (Na$^{+}$/H$_{3}$O$^{+}$ exchange) \cite{Hoppe2011}, then, the release of Ca$^{2+}$ and PO$^{3-}_{4}$ groups giving rise to a phosphocalcium-rich layer on the top of the silica-rich surface. The release of modifier ions (Na$^{+}$ and Ca$^{2+}$) from the glass affects the physiological balance of solution at the glass soft/hard tissue interface and modify the local pH \cite{ Lin2005}, in addition, Na$^{+}$ have an impact on the degradability of silicate network and plays principal control factor in ion release \cite{ Wallace1999, Murphy2009}. 

The 45S5 glass has been well investigated in the last decades. Gonz\'alez \textit{et al.} used Fourier transform Raman spectroscopy to highlight the structural role of the glass network modifiers in bioactive glasses and how it is related to the formation of the HCA \cite{Gonzlez2003}. Notingher \textit{et al.} with the help of Raman microspectroscopy, characterized the surface reactions of bioactive glasses when put in contact to simulated body fluid (SBF) \cite{Notingher2002}. One of the main limitations that prevent 45S5 bioglass from being an ideal biomaterial is those related to cytotoxicity, i.e., the high pH environment created by the high sodium content \cite{Fernandes2018}. Moreover, the high solubility and fast ions release might exceed those of the surrounding implantation environment before the formation of the new bone \cite{Rabiee2015} leading to \textit{in vivo} cytotoxic problem, particularly the dentine hypersensitivity as mentioned in literature \cite{Bakry2011}. In this regard, numerous investigations focused on improving the bioactivity, mechanical properties, anti-inflammatory, and antibacterial properties through compositional design \cite{Zhang2014, Gentleman2010, Atila2019a}. Computer simulations also played an important role in understanding the effect of the glass structure (short and medium-range order) on the bioactivity and other properties related to it \cite{Tilocca2009b, Tilocca2009a, Christie2011, Malavasi2013}. Enabling control of cytotoxicity by controlling the diffusion of the ions will affect the degradation mechanism for this type of biomaterials. Thus, understanding the effect of density on ionic diffusion behavior in 45S5 melts is a promising way to control ion release and limit the increase of pH in the surrounding tissue. As shown in several investigations for both experiments and computer simulations \cite{Mascaraque2017, Karki2010}, the hot densification has a crucial impact on macroscopic properties; this macroscopic change is correlated with structural features such as pairs bonding, angular distribution, and coordination numbers; which highlights the effect of the glass topology on controlling its properties.

We evaluate the impact of the melt density on the diffusion behavior of 45S5 bioactive melts (highly bioactive composition) using molecular dynamics simulations to understand how the density affects ions mobility and how this latter is correlated to the structural changes. A similar trend has been observed for all elements for either diffusion constant or activation energy, but the response was notable in modifiers, especially in Na cations.
In the following, the method followed to obtain the 45S5 melts is presented in Sec. \ref{Methods}. Section. \ref{Results}, presents the results where we show the effect of the density on the diffusion coefficients and activation energy in soda-lime phosphosilicate melts. Discussion of the results correlating between diffusion behavior, density, and structure is given in Sec. \ref{Discussion}. Finally, concluding remarks are given in Sec. \ref{Conclusion}.

\section{\label{Methods}Computational details}
The effect of the density on the dynamical behavior of 45S5 was studied using classical molecular dynamics. All simulations were performed using LAMMPS code \cite{Plimpton1995}. We rely on the well-established potential by Pedone \textit{et al.} \cite{Pedone2006} to model the interatomic interactions. This potential has been proven to gives a realistic agreement with available experimental data \cite{Atila2022a, Deng2019, Atila2019b, Atila2020b, Atila2020c, Atila2020a, Ghardi2019}. Potential parameters and partial charges are found in Ref. \cite{Pedone2006}. Long-range interactions were evaluated via the Ewald summation method, with a real-space cutoff of 12.0 \AA\ and precision of $10^{-5}$. The short-range interactions cutoff distance was chosen to be 5.5 \AA\ \cite{Pedone2006}. The simulations were run in the canonical ensemble (NVT) using the Nosé-Hoover thermostat. Velocity-Verlet algorithm with an integration time step of 1 fs was used to integrate the equations of motion. 

All systems consist of 4275 atoms placed randomly in a periodic cubic simulation box according to the 45S5 nominal molar composition (CaO)$_{26.9}$--(Na$_2$O)$_{24.4}$--(SiO$_2$)$_{46.1}$--(P$_2$O$_5$)$_{2.6}$ and avoiding any unrealistic overlap between atoms. The box length was changed to get different densities ranging from 2.32 g/cm$^3$ to 5.71 g/cm$^3$. First, we equilibrated the systems at a high temperature (T = 5000 K) for 500 ps. This step is needed to ensure that each system loses the memory of its initial configuration. After that, the systems were subsequently quenched linearly from the liquid temperature (T = 5000 K) to the room temperature (T = 300 K) with a cooling rate of $10^{12}$ K/s while keeping the volume fixed. During the cooling, configurations were extracted at a temperature T (T = 3500, 3400, 3300, 3200, 3100, 3000) and used as starting configurations for the diffusion simulations for each temperature. Then, the diffusion calculations were performed in the NVT ensemble for 2 ns. At 300 K, the glass was further equilibrated in the NVT ensemble for 1.5 ns. The structural and elastic properties were found to be in realistic agreement with the experimental data as discussed in the Ref. \cite{Ouldhnini2021a}. 

\section{\label{Results}Results}
The mean squared displacements (MSD) were calculated at different temperatures (T = 3000 - 3500 K),
\begin{equation}
    MSD = \left< |r(t) - r(0)|^2 \right>
\end{equation}
where $r(0)$ is the initial position at $t$ = 0 and $r(t)$ is the position at time t. The diffusion coefficient D was obtained using Einstein's equation,
\begin{equation}
    D = \lim_{t \to +\infty} \frac{MSD}{6t}
\end{equation}
and it was averaged over the last 200 ps of each run. The length of the simulation time used in this work is long enough for the MSD to be in the linear regime, thus giving good estimate of the diffusion coefficients.
\begin{figure*}[ht!]
\centering
\includegraphics[width=\textwidth]{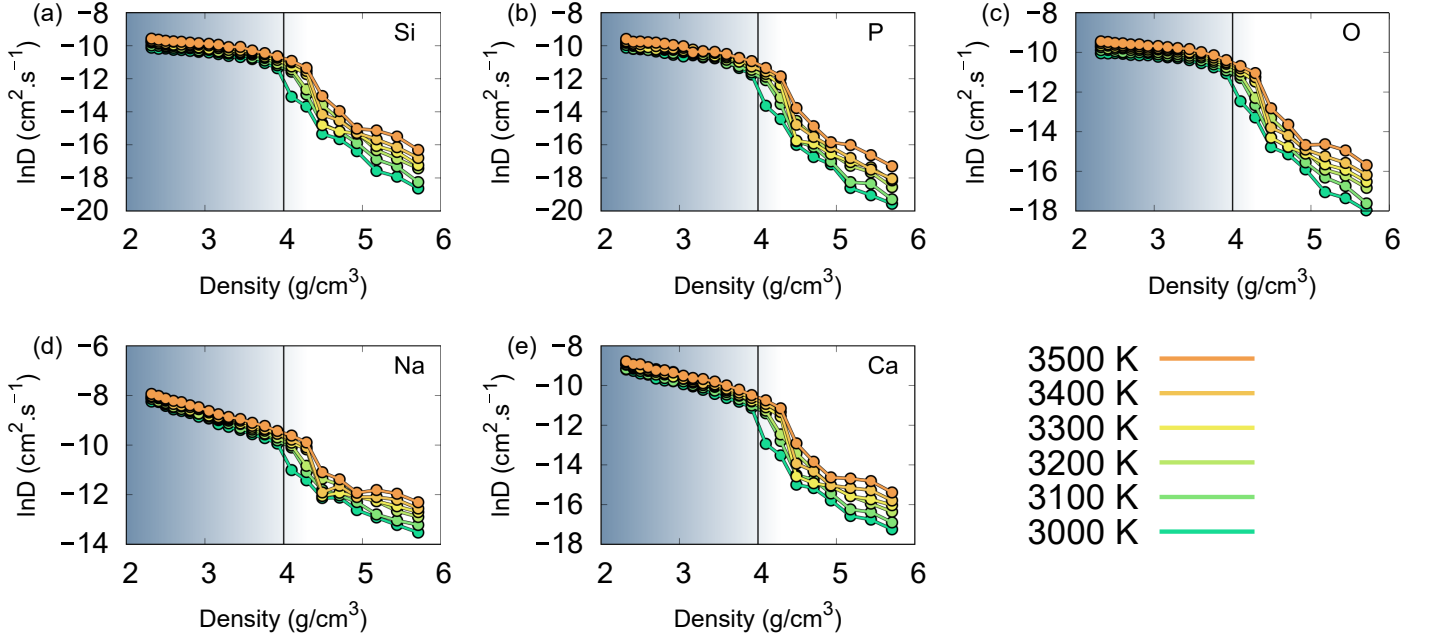}
\caption{Diffusion constant calculated at different temperatures for different densities. The blue shaded region indicates an NC less than 3, and the vertical line shows the density at which the NC is equal to 3. Lines are a guide for the eyes. Error bars are smaller than the symbol size.}
\label{Ddensity}
\end{figure*}

\begin{figure}[ht!]
\centering
\includegraphics[width=\columnwidth]{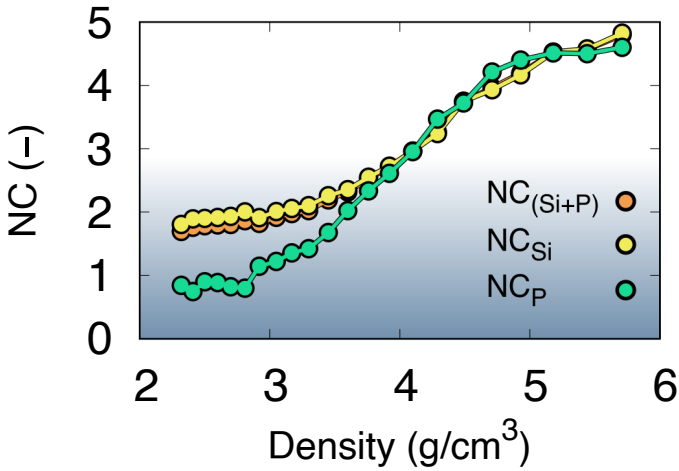}
\caption{Total Network connectivity and network connectivity of Si and P atoms as a function of the density at 3000 K. The blue shaded region indicates an NC less than 3. The symbols represent the simulated data points, and the lines are a guide to the eye; error bars are smaller than the symbol size, and they are omitted from the plot.}
\label{NCdensity}
\end{figure}

Figure~\ref{Ddensity} shows the diffusion coefficients as a function of density at various temperatures for all elements in the 45S5. The blue shaded region indicates a network connectivity NC less than 3, and the vertical line shows the density at which the NC is equal to 3 (see Fig.~\ref{NCdensity} for more details on NC). The diffusion of all elements decreases with increasing density, which is consistent with the diffusion behavior observed for Si and O ions in silica melt under pressure \cite{Karki2010} and with the data reported for magnesium silicate liquid under pressure \cite{Ghosh2011}. Three main diffusion regimes are observed, as indicated by the change of the slop in the variation of the diffusion coefficient with increasing density. These regimes are identified by two densities around 3.8 g/cm$^3$ and 4.8 g/cm$^3$, respectively. In addition to that, these density values are shifted to lower densities with decreasing temperatures. During the initial diffusion regime at low densities, the diffusion of all elements decreases slightly with increasing density. The diffusion of Na and Ca cations was more affected by the density increase than the glass former elements. In the intermediate diffusion regime located in the range between 3.8 and 4.8 g/cm$^3$, a sharp decrease of the diffusion coefficients curves for all elements appears for all temperatures, indicating a transition region. The last diffusion regime appears at high densities, in which diffusion continues its decrease with a slope that is lower than that of the intermediate regime and closer to that of the initial regime, and the isotherms diverge from each other with increasing density.

\begin{figure}[ht!]
\centering
\includegraphics[width=\columnwidth]{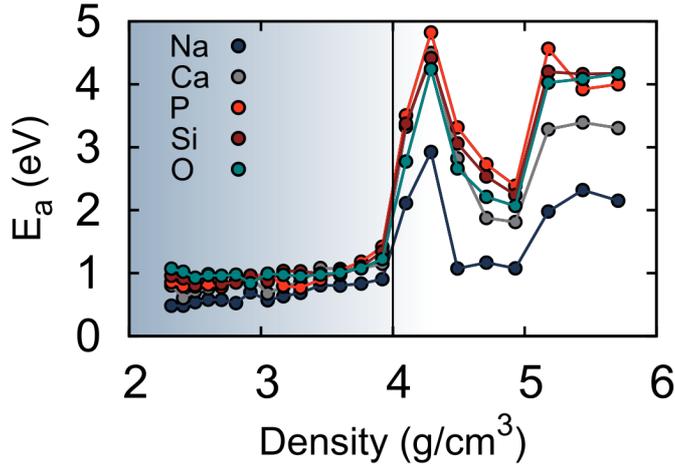}
\caption{Activation energies E$_a$ of Na, Ca, Si, P, and O as a function of the density. The blue shaded region indicates an NC less than 3, and the vertical line shows the density at which the NC is equal to 3. The symbols represent the simulated data points, and the lines are a guide to the eye; error bars are smaller than the symbol size, and they are omitted from the plot.}
\label{Eadensity}
\end{figure}

The lnD vs 1000/T plot follows an Arrhenius behavior that is depicted by the following equation,
\begin{equation}
lnD = lnD_0 - \frac{\Delta E_a}{k_BT}
\label{Arrheniuseq}
\end{equation}
where E$_a$ is the activation energy, D$_0$ is the pre-exponential factor, k$_B$ is the Boltzmann constant, and T is temperature \cite{Atila2020b}. From Eq.~\ref{Arrheniuseq} we can extract the activation energy that is the slope of a linear fitting between lnD and 1000/T. Figure~\ref{Eadensity} depicts the activation energy change as a function of density for all elements. The activation energies at ambient density are in good agreement with the data in the literature \cite{Xiang2011, Lu2018, Du2012}. We observe that all activation energies slightly increase for densities lower than 3.8 g/cm$^3$ and vary significantly at densities higher than 3.8 g/cm$^3$. Furthermore, the activation energies plot can also be divided into three regimes similar to the diffusion coefficients. For the first regime (densities lower than 3.8 g/cm$^3$), it is clear that E$_a$ curves are close to each other and nearly constant for O, Si, and P, while an increase has been observed for Na and Ca cations. In the second regime (densities between 3.8 g/cm$^3$ and 4.5 g/cm$^3$), all E$_a$ curves show a significant increase and reach a maximum at a density close to 4.3g/cm$^3$, then decrease and reach a minimum that is larger than that corresponding to the first regime. The activation energies increase again in the last regime (densities higher than 4.5 g/cm$^3$).

The dependence of the self-diffusion coefficient on the density shows that the pressure strongly affects the dynamics of the 45S5 melts with a nonlinear trend which implies that the diffusion coefficients do not follow the Arrhenian law for their pressure dependence. According to this law, the diffusion isotherms should be straight lines in the logarithmic plots, and their slopes should decrease with increasing temperature. The predicted complex P-T behavior of the diffusion coefficients can be described using a modified Arrhenius law by Ghosh \textit{et al.} \cite{Ghosh2011} as shown in Eq. \ref{arhpres}, 

\begin{equation}\label{arhpres}
D_{\alpha }=D_{0\alpha }\exp\left [  -\frac{E_{a}+PV_{\alpha}(P,T)}{k_{B}T} \right ],
\end{equation}
where $k_{B}$ is the Boltzmann constant, $D_{0\alpha}$ and $E_{\alpha}$ are the pre-exponential and the averaged activation energy for an atom type $\alpha$, respectively, and their values were taken from zero-pressure Arrhenian plots, $V_{\alpha}(P,T)$ is the activation volume and equal to $V_{0}+V_{1}T+P(V_{2}+V_{3}T+V_{4}/T)$, these fitting coefficients represent the pressure, temperature and cross derivatives of the activation volumes and are summarized in Table~\ref{Parameters}.

\begin{table*}[ht!]
\centering
\caption{Activation volumes and their derivatives for the modified Arrhenian models (Eq. \ref{arhpres}) of the self-diffusion of all atom types in the simulated 45S5 glasses.}
\centering
\begin{tabular}{cccccc} 
\hline
\hline
Parameters & V$_{0}$(cm$^3$mol$^{-1}$) & V$_{1}$(cm$^3$mol$^{-1}$K$^{-1}$) & V$_{2}$(cm$^3$mol$^{-1}$GPa$^{-1}$) & V$_{3}$(cm$^3$mol$^{-1}$K$^{-1}$GPa$^{-1}$) & V$_{4}$(cm$^3$mol$^{-1}$K GPa$^{-1}$)  \\ 
\hline
$D_{Na}$ & 0.35 ($\pm$0.03) & -9.04$\times$10$^{-5}$ ($\pm$8$\times 10^{-6}$)& -0.0058 ($\pm$3$\times 10^{-4}$)& 1.105$\times 10^{-6}$ ($\pm$6$\times 10^{-8}$)& 6.65 ($\pm$1) \\
$D_{Ca}$ & 0.49 ($\pm$0.04)& -13.33$\times$10$^{-5}$ ($\pm$1$\times 10^{-6}$)& -0.1093 ($\pm$5$\times 10^{-4}$)& 2.145$\times 10^{-6}$ ($\pm$1$\times 10^{-7}$)& 12.43 ($\pm$3) \\
$D_{Si}$ & 0.30 ($\pm$0.02)& -8.952$\times$10$^{-5}$ ($\pm$6$\times 10^{-6}$)& -0.0156 ($\pm$9$\times 10^{-4}$)& 2.873$\times 10^{-6}$ ($\pm$1$\times 10^{-7}$)& 21.44 ($\pm$2) \\ 
$D_{P}$ & 0.28 ($\pm$0.03)& -8.407$\times$10$^{-5}$ ($\pm$8$\times 10^{-6}$)& -0.0155 ($\pm$2$\times 10^{-3}$)& 2.816$\times 10^{-6}$ ($\pm$3$\times 10^{-7}$)& 21.98 ($\pm$3) \\
$D_O$ & 0.26 ($\pm$0.02)& -8.031$\times$10$^{-5}$ ($\pm$5$\times 10^{-6}$)& -0.0170 ($\pm$8$\times 10^{-4}$)& 3.032$\times 10^{-6}$ ($\pm$2$\times 10^{-7}$)& 24.81 ($\pm$2) \\
\hline
\hline
\end{tabular}
\label{Parameters}
\end{table*}

The activation volumes of all elements function the pressure and temperature. Thus, by using the values of the cross derivatives provided in Tab.~\ref{Parameters}, we can qualitatively predict the diffusion behavior at different temperature and pressure conditions. The activation volume $V_{\alpha}(P,T)$ of Na and Ca at zero pressure and 3000 K are the highest ($V_{Na}=0.077$ cm$^3$ mol$^{-1}$, $V_{Ca}=0.09$ cm$^3$ mol$^{-1}$) compared to Si, P and O elements ($V_{Si}=0.034$ cm$^3$ mol$^{-1}$, $V_{P}=0.034$ cm$^3$ mol$^{-1}$, and $V_{O}=0.022$ cm$^3$ mol$^{-1}$), these values increase as the liquid is compressed for all the temperatures. We find that our prediction falls within the diffusion behavior at 3000 K \cite{Ghosh2011} (concave isotherm in a logarithmic plot), in which a nonlinear diffusion for Si, P, and O is observed and a weaker nonlinear one for the modifiers.

\begin{figure*}[ht!]
\centering
\includegraphics[width=\textwidth]{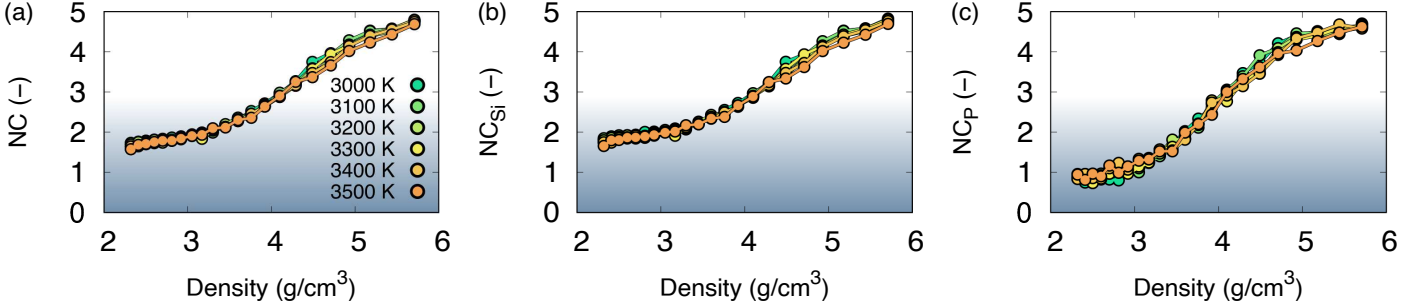}
\caption{Total Network connectivity (a) and partial network connectivity of Si (b) and P (c) atoms as a function of the density at different temperatures. The blue shaded region indicates an NC less than 3. The symbols represent the simulated data points, and the lines are a guide to the eye; error bars are smaller than the symbol size, and they are omitted from the plot.}
\label{NCTdensity}
\end{figure*}

The network connectivity (NC) is regarded as a good structural descriptor of glass bioactivity \cite{Edn2011, Jones2013}. The NC was calculated using Eq.~\ref{NCeq} 
\begin{equation}
\label{NCeq}
NC = \sum_{n=0}^n n x_n
\end{equation}
where x$_n$ is the fraction of the Q$^n$ (defined as the average number of bridging oxygen (BO) atoms bound to a network-forming cation) (with n = 0, 1, 2, 3, 4, 5, or 6). The NC$_{Si}$, NC$_P$, and total NC as a function of the density at 3000 K are displayed in Fig.~\ref{NCdensity}. The NC increases with increasing density and shows qualitatively three regimes. The plots of the total NC and NC$_{Si}$ are almost superposed and have a nearly constant value around 2 in the first regime, while that of P is much lower. A significant increase in the NC characterizes the intermediate regime. In the last regime, at higher densities, the NC reaches values around 4.8 and shows a weak variation with increasing density. The temperature and density dependence of NC is shown in Fig.~\ref{NCTdensity} and describes a similar behavior as the one in Fig.~\ref{NCdensity} with negligible temperature dependence.

\section{\label{Discussion}Discussion}
As shown in Fig.~\ref{Ddensity} and Fig.~\ref{Eadensity}, as temperature decreases, the system dynamics slows down significantly, which is due to a reduction in the kinetic energy, making atoms unable to jump over higher energy barriers and/ or when the melt density increases the volume decreases which leads to increasing the energy barriers and induces a structural change (i.e., cage) that limits the atom movement. The dynamical behavior of the 45S5 glasses can be explained by the changes of the local structure and the polymerization of the network in the form of an increase of bridging oxygen (BO) and the appearance of oxygen tri-clusters (O$_3$, TBO) and a decrease of non-bridging oxygen (NBO) making the structure more compact. 

\begin{figure}[ht!]
\centering
\includegraphics[width=\columnwidth]{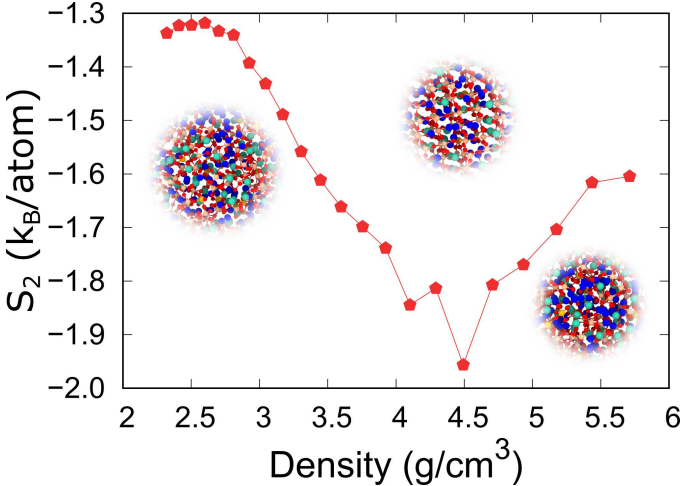}
\caption{The two-body excess entropy (S$_2$) in the system normalized by the total number of atoms (N) in the system as a function of the density at 3000 K. The inset are snapshots taken at different densities (2.7, 4.1, and 5.7 g/cm$^3$). The symbols represent the simulated data points, the lines are a guide to the eye, and error bars are smaller than the symbol size.}
\label{Entropydensity}
\end{figure}

\begin{figure*}[ht!]
\centering
\includegraphics[width=\textwidth]{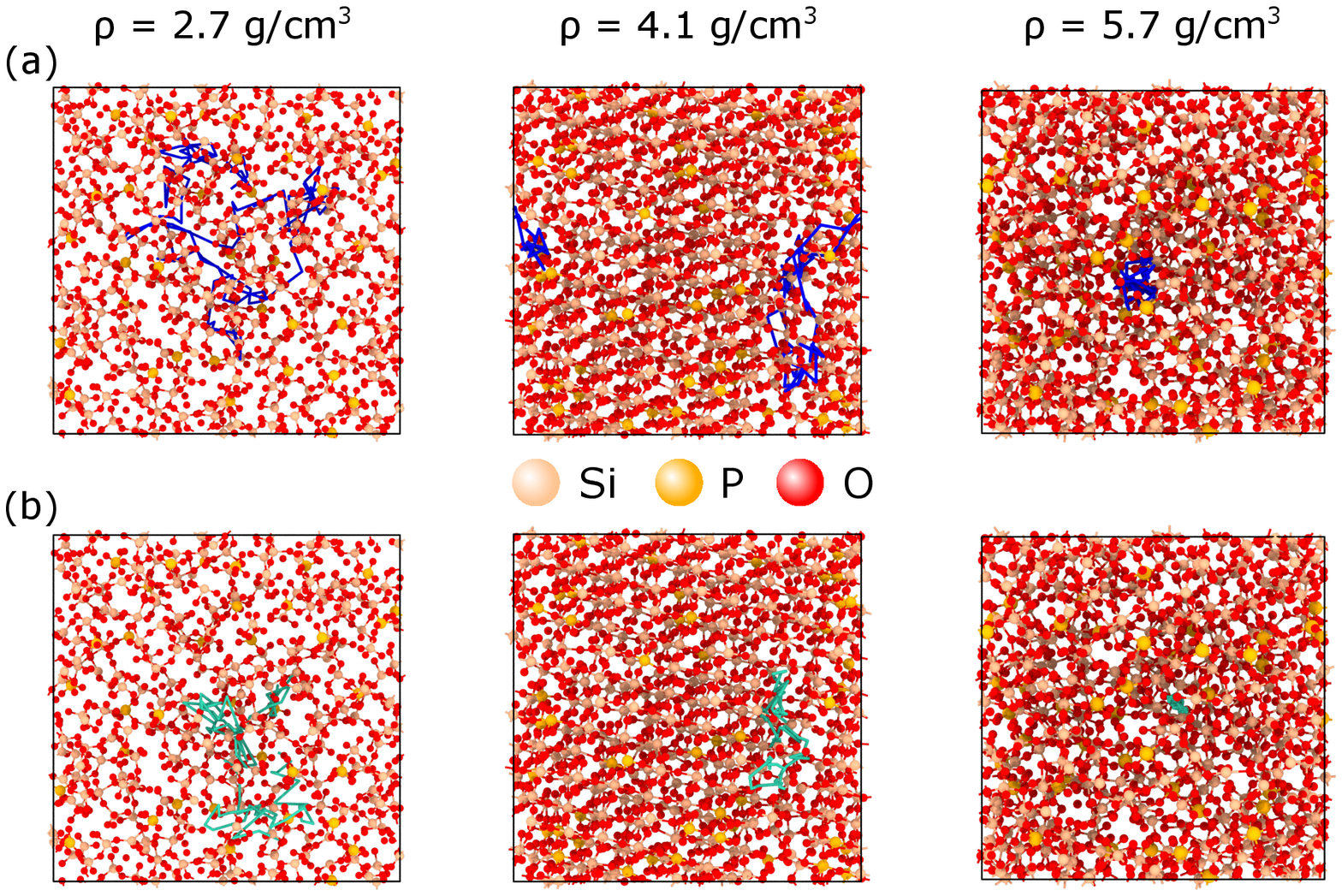}
\caption{Snapshots are showing trajectories of selected sodium (a) and calcium (b) atoms during the last 100 ps of an NVT run at a temperature of 3000 K and for selected densities (2.7, 4.1, and 5.7 g/cm$^3$). Blue and green lines highlight the trajectories of selected sodium and calcium atoms, respectively. Only Si, P, and O atoms are shown, Na and Ca atoms are removed for clarity.}
\label{Snapshots}
\end{figure*}

In a low-density regime, the diffusion coefficient and the activation energy for Si, P, and O slowly decrease and increase, respectively, as the density increases. This is because the modifiers move freely in the open space available, and this is mainly because Si and P atoms are enclosed in tetrahedral units (four-fold connection) with strong bonding to O atoms, in contrast to the weak bonding of Na and Ca with O atoms. This difference is confirmed because modifiers have a larger activation volume. In addition to the activation volumes, the change in the diffusivity of the atoms can be explained through the pair excess entropy, S$_2$, which approximates the configurational entropy \cite{Atila2020a}. Higher values of S$_2$ indicate more disorder in the structure. The pair excess entropy initially remained almost constant, then decreased with increasing density, indicating that the glass structure becomes more ordered at higher densities.
Moreover, it increased again after reaching a minimum at a density of 4.5 g/cm$^3$. The change of S$_2$ is consistent with the change in the diffusion and highlights a disorder-order-disorder pressure-driven transition, as indicated by the snapshots in Fig.~\ref{Entropydensity}(inset) and Fig.~\ref{Snapshots}. The increase of the network connectivity resulted in a decrease of the diffusion of Na and Ca, as shown by the shorter diffusion paths at higher densities (See Fig.~\ref{Snapshots}). Moreover, the change in the slope of the diffusion shown in the intermediate density regime (See Fig.~\ref{Ddensity}) is mainly due to a partial ordering of the network at high temperatures, which is seen in Fig.~\ref{Snapshots} for the density around 4.1 g/cm$^3$. This behavior has been previously observed in metallic glasses \cite{Atila2020b, Hu2017}. The Na atoms diffuse faster and explore larger space than Ca atoms even at higher densities, which is consistent with previously reported results \cite{Atila2020a}. Similar behavior was predicted for the Ca diffusion in anorthite liquid \cite{Karki2011}. 

In the intermediate density regime, the decrease of the diffusion is found to be more significant; in this regime, the atoms need more energy to jump from their cages and diffuse, which is due to the observed partial crystallization or the formation of an ordered network made out of Si/PO$_6$ structural units. The decrease of the activation energies noticed at a density around 4.5 g/cm$^3$ is due to atoms mostly diffusing in an ordered system of linked octahedra. The final increase of E$_a$ is mainly related to the system shrinking these octahedra, leading to a more locked network hindering the atom's dynamic. Thus, compression, to some extent, improves the mechanical properties and chemical durability of the glasses. The effect of permanent densification on the glass topology and its impact on the dissolution mechanisms was studied by Mascaraque \textit{et al.}, \cite{Mascaraque2017}, where they reported that the chemical durability increases with increasing density showing the impact of the glass topology on dissolution behavior. This is in line with the presently reported results, as the dissolution kinetics is strongly dependent on the dynamic behavior of the sample. Moreover, several experiments are consistent with our results \cite{Zhang1991a, Zhang1991b}. Our findings present a modest contribution toward controlling the transport properties of bioactive glasses, which is necessary for further development and rational design of bioactive glasses with adapted properties.

\section{\label{Conclusion}Conclusion}
Molecular dynamics simulations have been performed to study the effect of the density on the diffusion behaviors of 45S5 sample in a range of temperatures between 3000 and 3500 K. We showed that the density influences the mobility of the ions, which is seen by a decrease of the diffusion constant and an increase of the energy barriers for self-diffusion. Moreover, a transition from amorphous to an ordered structure and back to amorphous was observed and verified by decreasing the pair excess entropy. The observed changes are well explained and correlated with the coordination changes and the repolymerization of the network with increasing density. The importance of the diffusion behavior of bioactive glass in determining the dissolution kinetics and, in turn, the bioactivity, the results presented in this paper may provide theoretical guidance and pave a new route toward the design of functional bioactive glasses with controlled properties.

\section*{Conflicts of interest}
There are no conflicts to declare.

\section*{Acknowledgments}
A. Atila thank the German Research Foundation (DFG) for financial support through the priority program SPP 1594 – Topological Engineering of Ultra-Strong Glasses. The authors gratefully acknowledge the computing resources provided by the Erlangen Regional Computing Center (RRZE).

\bibliographystyle{unsrt}
\bibliography{biblio}

\end{document}